\documentclass[showpacs,preprintnumbers,amsmath,amssymb,eqsecnum,nofootinbib]{revtex4}

\def\bv{{\boldsymbol \varphi}}
\def\bn{{\mbox{\boldmath$n$}}}

\def\bPhi{{\mbox{\boldmath$\Phi$}}}
\def\bphi{{\mbox{\boldmath$\phi$}}}
\def\bP{{\mbox{\boldmath$P$}}}
\def\etain{{-\infty}}
\def\bbvp{{\bv\hspace{-.72em}\bv}}
\def\bbp{{\bphi\hspace{-.70em}\bphi\hspace{-.71em}\bphi}}
\def\bbn{{\bn\hspace{-.70em}\bn\hspace{-.71em}\bn}}

\usepackage{amssymb}
\usepackage{bm}
\usepackage[dvips]{graphicx,color}

\begin{document}
\thispagestyle{empty}

\vspace*{-25mm}
\begin{flushright}
\baselineskip14pt
KAIST-TH/2006-06\\
YITP-06-18 \\
KUNS-2022 
\end{flushright}

\title{
Wronskian Formulation of the Spectrum of Curvature Perturbations}
\author{
Shuichiro Yokoyama$^1$,
Takahiro Tanaka$^1$,
Misao Sasaki$^2$,\\
and Ewan D. Stewart$^{3,4,5}$
}
\affiliation{
$^1$Department of Physics, Kyoto University, Kyoto, Japan\\
$^2$YITP, Kyoto University, Kyoto, Japan\\
$^3$Department of Physics, KAIST, Daejeon, Republic of Korea\\
$^4$Department of Physics and Astronomy, University of Canterbury,
Christchurch, New Zealand\\
$^5$CITA, University of Toronto, Toronto, Canada\\
}
\begin{abstract}
We present a new formulation for the evaluation of
the primordial spectrum of curvature perturbations generated 
during inflation, using the fact that the Wronskian of the
scalar field perturbation equation is constant.
In the literature, there are many works
on the same issue focusing on a few specific aspects or 
effects. Here we deal with the general multi-component
scalar field, and show that our new formalism gives 
a method to evaluate the final amplitude of the 
curvature perturbation systematically and economically. 
The advantage of the new method is that one only has to
solve a single mode of the scalar field perturbation
equation backward in time from the end of inflation to the stage
at which the perturbation is within the Hubble horizon, at
which the initial values of the scalar field perturbations are given.
We also clarify the relation of the
new method with the new $\delta N$ formalism recently developed in
Ref.~\cite{ref4}.
\end{abstract}

\maketitle


\section{Introduction}

Inflation provides an elegant mechanism to solve 
shortcomings of the standard Big Bang Model, for example, 
the flatness and horizon problems. 
Moreover, in the inflationary universe, curvature perturbations, 
which seed the structure formation of the universe, 
are generated from vacuum fluctuations of the scalar field. 
One can test various models of inflation, by 
comparing the theoretical prediction for 
the spectrum of curvature perturbations with observations. 

In single-field slow-roll inflation models, 
curvature perturbations evaluated on comoving 
hypersurfaces, ${\cal R}_{\rm c}$, are constant on 
super-horizon scales, and their spectrum is given by~
\begin{equation}
{\cal P}_{{\cal R}_{\rm c}}(k) \sim 
\left(\frac{H^2}{2\pi\dot{\phi}}\right)^2_{aH = k},
\label{standard}
\end{equation}
where $H$ is the Hubble parameter, $\dot{\phi}$ is the derivative of 
the inflaton $\phi$ with respect to the cosmological time, 
and the subscript ``$aH = k$'' means that the expression is to be evaluated 
at the horizon crossing time, $t = t_k$.
This equation indicates that the spectrum of the primordial curvature
perturbations is almost scale-invariant in slow-roll inflation  
models composed of a single scalar field. 

However, 
in making realistic models of inflation 
based on supersymmetry and supergravity theories, it seems 
more natural to consider models composed of 
a multi-component scalar field, in which 
the slow-roll conditions are also possibly violated~\cite{Kadota:2003fs}. 
Obviously, Eq.~(\ref{standard}) is not sufficient to evaluate 
the spectrum of the curvature perturbations generated in such models.
In order to identify necessary conditions for viable models of inflation
among various proposed
possibilities, any simple and sufficiently accurate formula 
for the spectrum is useful, especially if the formula is 
applicable to a wide class of models. 

There are several works aiming at generalizing 
the slow-roll formula~(\ref{standard})  
in the case of a single scalar field.
For example, 
Leach, Sasaki, Wands and Liddle~\cite{ref2} have analyzed the evolution of 
curvature perturbations on superhorizon scales (the long-wavelength
approximation). They have found that the curvature perturbations can be
enhanced when $z \equiv a\dot{\phi}/H$
decreases after the horizon crossing time $t_k$. 
This situation can happen 
if the slow-roll conditions are violated for $t>t_k$. 
On the other hand, Stewart~\cite{ref3} has generalized the standard 
formula (\ref{standard}), removing 
the extra assumption that slow-roll parameters are
approximately scale invariant, and this generalization is 
called the general slow-roll approximation. 
Under the conditions which lead to the standard slow-roll 
formula~(\ref{standard}), the running 
of spectrum is automatically suppressed as 
$\left|dn_{\cal R_{\rm c}}/d\ln k\right|\ll\left|n_{\cal R_{\rm c}}-1\right|$,
while under the general slow-roll conditions,
$\left| d n_{\cal R_{\rm c}}/d \ln k \right|$ 
can be as large as
$|n_{\cal R_{\rm c}}-1|$. 
These conditions are consistent with recent observational data~\cite{WMAP,Spergel:2006hy}.

For multi-component inflation, 
evaluating the curvature perturbations is a more difficult issue 
in comparison with single-component inflation
\cite{SS,ST,multi,Wands:2000dp,Nakamura:1996da,Salopek:1988qh,Bartolo:2001rt,Hwang:1999gf,
Starobinsky:1994mh,Polarski:1994rz,Garcia-Bellido:1996ke,Langlois:1999dw,Kadota:2003fs}.
Generally, in the case of a $D$-component field, the evolution equations
for the perturbations are $D$ coupled second order differential equations.
It is a heavy task to solve the full set of equations. 
Moreover, for multi-component inflation,
the curvature perturbations on comoving slices, $\mathcal{R}_{c}$,
still varies in time on super-horizon scales 
until the neighboring 
trajectories of the background homogeneous universe in the phase space 
practically converge to a single one.
It is after the convergence of the trajectories 
that $\mathcal{R}_{c}$ becomes constant in time~\cite{SS,ST}.

However, in the estimation of the spectrum, what we need is only
the final value of the curvature perturbation $\mathcal{R}_{c}(\eta_{f})$.
Here, $\eta$ is a conformal time given by $\eta = \int dt/a$, and 
$\eta_{\rm\, f}$ is taken sufficiently late after the convergence of the background
trajectories. So, if we can identify the part of perturbations
that contributes to $\mathcal{R}_{c}(\eta_{f})$,
we may solve only that part to obtain its value
without solving the full set of perturbation equations.
The $\delta N$ formalism~\cite{SS,ST} was based on this idea, and
we have developed a new $\delta N$ formalism~\cite{ref4}.  
This formulation was developed based on the fact
that the perturbation of $e-$folding number stays roughly constant 
in the flat slicing on super-horizon scales. 
We derived a perturbation equation for a single projected 
component of scalar field corresponding to 
the perturbation of $e-$folding number. 
The derivation of the new $\delta N$ 
formalism is rather intuitive in this sense, 
which helps to make its physical meaning rather transparent. 
As a drawback, however, the systematical derivation of the formula 
is not so simple. 
Moreover, in Ref.~\cite{ref4} we only applied our new formalism to cases 
without the possible super-horizon contributions pointed out in Ref.~\cite{ref2}, 
postponing this generalized application to Ref.~\cite{future}.
Here we propose an alternative slightly different framework 
to evaluate the evolution of perturbations of a 
multi-component scalar field. 
The method we will discuss in this paper takes full advantage 
of the fact that the Wronskian stays constant. 

This paper is organized as follows.
In section \ref{secdeltaN}, we briefly explain the 
perturbation equations for multi-component 
scalar field in the flat slicing. 
In section \ref{Wronskian},
we introduce a Wronskian between two solutions of 
perturbation equations, and explain how we can use it in 
evaluating the curvature perturbation at a later time $\eta=\eta_f$. 
As far as we are concerned with the evolution 
during the phase dominated by the scalar field, 
what we have to do turns out to be just solving a single mode backward
in time with appropriate boundary conditions imposed at
$\eta=\eta_{\rm\, f}$. 
In section \ref{extended}, we use the Wronskian to 
derive more explicit analytic expressions for the spectrum of
perturbations. 
We consider an extension of the general slow-roll 
formula developed in Ref.~\cite{ref3,ref4}, 
combining it with the long wavelength approximation~\cite{ref2}, 
which suites for describing the late time evolution~\cite{future}. 
We assume that the general slow-roll approximation is valid until a
time $\eta_*$ later than the horizon crossing time $\eta_k$,  
and we use the long-wavelength approximation for 
the succeeding evolution until the end of inflation, $\eta_{f}$. 
To combine the two solutions obtained by using 
different approximation schemes in the matching region 
where both approximations are valid, we find that 
the constancy of the Wronskian is again very useful. 
The derived new analytic formula applies for a wider class of
multi-component inflation models. 
In section \ref{secsingle}, we reduce the obtained formula to 
the single-field case. In this case, 
the formula for the spectrum can be expressed more explicitly. 
Using our new formulation, it is also easy to see that 
the amplitude of perturbation stays regular even if the time 
derivative of $\phi$ vanishes, as was discussed in Ref.~\cite{Seto}. 
In section \ref{summary}, we present a brief summary.

\section{Perturbation equations for multi-component scalar field}
\label{secdeltaN}

We consider a $D$-component scalar field whose action is given by
\begin{equation}
S_{{\rm matter}} = - \int d^4x\sqrt{-g}
 \left[\frac{1}{2}g^{\mu\nu}\delta_{\alpha\beta}\partial_{\mu}\phi^{\alpha}\partial_{\nu}\phi^{\beta}+ V(\phi)\right]~,
\end{equation}
where, for simplicity, we have chosen to work with the flat metric
$\delta_{\alpha\beta}$ for the scalar field space, because 
the equations will be more involved for the general field space
metric and hence the essence of our new formulation may be obscured,
though the extension to the general metric case is straightforward.
The background equations are
\begin{eqnarray}
3{\cal H}^2 = \frac{1}{2}\delta_{\alpha\beta}{\phi^{\alpha}}'{\phi^{\beta}}'+a^2V(\phi)~,\\
{\phi^\alpha}'' + 2{\cal H}{\phi^\alpha}'+a^2\delta^{\alpha \beta}V_{,\phi^\beta} = 0~,
\end{eqnarray}
where ${\cal H} \equiv a'/a$, 
the prime represents a derivative with respect to 
the conformal time $\eta$, and 
$V_{,\phi^\alpha} \equiv \partial V/\partial \phi^{\alpha}$.

We consider only scalar type perturbations. 
Then the perturbed metric is given by~\cite{KS} 
\begin{equation}
ds^2 = a^2\left\{-(1+2A Y)d\eta^2 - 2B Y_jd\eta dx^j + \left[(1 + 2H_{L} Y)\delta_{ij}+2H_{T} Y_{ij}\right] dx^i dx^j \right\}~,
\end{equation}
where $Y$ is the spatial scalar harmonic function with the eigenvalue $k^2$, $Y_j \equiv -k^{-1}\nabla_j Y $, and 
$Y_{ij} \equiv \left(k^{-2}\nabla_i \nabla_j+\frac{1}{3}\delta_{ij}\right)Y$. 

We decompose the scalar field to the background and perturbations as 
\begin{equation}
\tilde{\phi} = \phi(\eta) + \delta \phi(x^\mu)Y~, 
\end{equation}
and define a gauge invariant variable
\begin{equation}
\delta \phi_F^\alpha \equiv \delta \phi^\alpha - \frac{{\phi^\alpha}'}{{\cal H}}{\cal
 R}~,
\label{deltaphiF}
\end{equation}
which represents 
the scalar field perturbation on the flat slicing.
Here
\begin{eqnarray}
{\cal R} \equiv H_L + \frac{1}{3}H_T~,
\end{eqnarray}
represents the intrinsic curvature perturbation of 
the constant time hypersurfaces. 
The evolution equation for $\delta \phi^\alpha_F$ is known to be given by~\cite{SS}
\begin{eqnarray}
\left\{\partial_\eta^2  - \left({\cal H}^2 + {\cal H}'\right) + k^2\right\} \varphi^\alpha 
= \left\{ \frac{1}{a^2}\left(\frac{a^2{\phi^\alpha}'{\phi^\gamma}'}{{\cal H}}\right)' \delta_{\gamma\beta}
 - a^2 \delta^{\alpha\gamma}V_{,\phi^\gamma\phi^\beta}\right\} \varphi^\beta~,
\label{flatEq}
\end{eqnarray}
where 
\begin{eqnarray}
\varphi^\alpha \equiv a\delta\phi^\alpha_F.
\end{eqnarray}

\section{Use of Wronskian}
\label{Wronskian}

When we calculate the curvature perturbation 
spectrum, we only need
the final value of the curvature perturbation,
$\mathcal{R}_{c}(\eta_{f})$. 
We do not need to know the evolution of all components of 
the multi-component field.
In other words, if we can identify the part of the perturbations
that contributes to $\mathcal{R}_{c}(\eta_{f})$,
we may be able to solve only the necessary part 
without obtaining a full set of solutions of perturbation equations.

The equation (\ref{flatEq}) is an equation which 
formally takes the form of 
\begin{equation}
 \left[\partial_\eta^2+Q(\eta)+k^2\right]\varphi^\alpha = P^\alpha_{~\beta}(\eta) \varphi^\beta, 
\label{simplifiedequation} 
\end{equation}
with $P^\alpha_{~\beta}$ 
symmetric with respect to the indices $\alpha$ and $\beta$. 
The Wronskian 
\begin{equation}
W(\bbn,\bbvp)
   \equiv  \bbn' \cdot \bbvp 
        - \bbn \cdot \bbvp', 
        \label{WR}
\end{equation}
introduced in a usual manner is constant  
if $\bbn$ and $\bbvp$ are solutions of 
Eq.~(\ref{simplifiedequation}). 
Here we used vector notation $\bbn$ and $\bbvp$ to represent 
 $n^\alpha$ and $\varphi^\alpha$ for brevity 
and $\bbn \cdot \bbvp \equiv \delta_{\alpha\beta} n^{\alpha} \varphi^{\beta}$. 

As is known well, when the universe is dominated by a 
single matter component whose physical state is solely characterized by 
a single parameter, say, the energy density as in the case of  
a slowly rolling scalar field or a single component perfect fluid, 
the curvature perturbation on comoving hypersurfaces ${\cal R}_c$
is constant in time on super-horizon scales.
If the background trajectories
in the phase space of the multi-component scalar field 
well converge to a single one at a time, $\eta_{\rm con}$, before the scalar field
dominant phase ends, the system is effectively 
equivalent to a single scalar field for $\eta>\eta_{\rm con}$. 
In such cases the problem is relatively easy to formulate. 
In comoving gauge, $\delta_{\alpha\beta}{\phi^\alpha}' \delta \phi^{\beta}=0$ by definition.  
Contracting Eq.~(\ref{deltaphiF}) with ${\phi}_\alpha' = \delta_{\alpha\beta}{\phi^\beta}'$, we find 
\begin{equation}
 {\cal R}_c=-{{\cal H} \bbp'\cdot \bbvp\over a{\bbp'}^2}. 
\end{equation}
Hence, by defining $\bbn$ so that it satisfies the boundary 
conditions 
\begin{equation}
 \bbn' =-{{\cal H}\bbp'\over a{\bbp'}^2}, 
\qquad
 \bbn=0,
\label{basicbc}
\end{equation}
at $\eta=\eta_{\rm con}$, the above Wronskian agrees with ${\cal R}_c$: 
\begin{equation}
{\cal R}_c(\eta_{\rm con})=W(\bbn,\bbvp). 
\label{RceqW}
\end{equation}
Since the Wronskian is constant, it can be evaluated at any time. 
Therefore, once $\bbn$ is solved backward in time until the wavelength 
of the mode is well inside the horizon scale
with the appropriate boundary conditions, Eq.~(\ref{basicbc}), we do not have
to evolve $\bbvp$ at all in the forward direction. This 
will be advantageous 
in the case of a multi-component field. In order to solve  
$\bbvp$ in the forward direction, we need to solve all 
$2D$ independent modes, where $D$ is the number of species. 
In contrast, to evolve $\bbn$ backward in time, we only need 
to solve a single mode. Hence, the advantage of solving $\bbn$ 
instead of $\bbvp$ becomes prominent 
as the number of species increases. 
An extreme example is the models 
with extra-dimensions~\cite{RubakovKobayashi}. 
Since $D$ is infinite in the brane world models, 
solving for all $\bbvp$ is impossible. 
Note that the boundary condition~(\ref{basicbc}) is given solely
in terms of the background quantities. This implies that $\bbn$
is uniquely determined for arbitrary scalar field perturbations.

In general multi-component models, 
it may not be a good assumption to assume that the trajectories 
converge to a single one before the scalar dominant phase terminates. 
Once the convergence of trajectories occurs,  
the universe undergoes a universal evolution everywhere, 
and the only remaining modes of scalar-type perturbations are
adiabatic ones~\cite{SS}. 
If we do not consider the case with isocurvature perturbations
that persist until the present epoch, we can assume that 
the convergence of trajectories occurs at a certain time,   
but it need not happen during the scalar dominant phase. 
In multi-component models, the convergence of trajectories may occur 
in general after complete reheating~\cite{SS,ref4}. 
In such cases it is at this stage that 
$\mathcal{R}_{c}$ becomes constant in time on
super-horizon scales. 

Even when the trajectories in phase space have not converged, 
the $\delta N$ formalism~\cite{SS} 
simplifies the evaluation of the evolution of perturbations 
in the long wavelength limit. 
In the rest of this section 
the long wavelength limit also contains the meaning
that we neglect the shear of the constant time hypersurfaces, 
which rapidly decays on super-horizon scales~\cite{ST}.
Let us take $\eta_{\rm\, i}$ to be a certain time soon after the relevant
scale have become greater than the horizon scale
and $\eta_{\rm\, f}$ to be a certain time after the 
complete convergence of trajectories. 
The $\delta N$ formalism is based on the fact that 
the final value of the curvature perturbation 
$\mathcal{R}_{c}(\eta_{\rm\, f})$ is given by
$\delta {\cal N} (\eta_{\rm\, f},\eta_{\rm\, i})$, the
perturbation in the $e$-folding number between the initial
flat hypersurface at $\eta=\eta_{\rm\, i}$ and 
the final comoving hypersurface at
$\eta=\eta_{\rm\, f}$ in the long wavelength limit.


The evolution of $\delta{\cal N}$ in the long wavelength limit 
is expected to be described by the dynamics of the homogeneous 
universe in the following sense~\cite{SS,ST}. 
We define $N$ by the $e$-folding number necessary to reach 
the point in the phase space corresponding to $\eta_{\rm\, f}$ 
by solving the evolution of the 
homogeneous universe. 
If we choose $\eta_{\rm\, i}$ to be in the scalar dominant phase,
$N$ will be a function of $\bbp$ and $\bbp'$. 
Then, in the super-horizon scales, the perturbation in $N$ is given by
\begin{eqnarray}
\delta N
 &=& 
  \left[{\partial N \over \partial \bbp}\cdot\delta\bbp_F
 +{\partial N \over \partial 
  \left(d{\bbp}/d\tau\right)} \cdot \delta\left({d\bbp\over d\tau}\right)_F
\right]_{\eta_{\rm\, i}}\nonumber\\
    &=& \left\{{\partial N \over \partial \bbp}\cdot\left({\bbvp \over a}
\right) 
      + {\partial N \over \partial{\bbp}'} \cdot \left[\left({\bbvp \over a}
\right)'-\bbp'A_F \right]\right\}_{\eta_{\rm\, i}}~.
\end{eqnarray}
Here, the subscript $F$ denotes a quantity evaluated on flat
hypersurfaces and the subscript 
$\eta_{\rm\, i}$ denotes a quantity evaluated at
$\eta=\eta_{\rm\, i}$. The perturbed proper time $\tau$ satisfies  
$d\tau = a (1+A)d\eta$, and the perturbation of the lapse 
function on flat hypersurfaces $A_F$ is given by~\cite{ref4}
\begin{equation}
A_F={1 \over 2}{\bbp'\cdot \bbvp \over a{\cal H}} .
\end{equation}
Thus we obtain
\begin{eqnarray}
\delta N
&=&-W(\bbn_B,\bbvp)_{\eta_{\rm\, i}}, 
\label{deltaNeqW}
\end{eqnarray}
with
\begin{equation}
 \bbn_B'=-{1 \over a}{\partial N \over \partial \bbp}
+{{\cal H} \over a}{\partial N \over \partial \bbp'}
 +{1 \over 2a{\cal H}}
\left({\partial N \over \partial \bbp'}\cdot \bbp'\right)\bbp', 
\qquad
 \bbn_B={1 \over a}{\partial N \over \partial \bbp'}. 
\label{deltaNboundary}
\end{equation}

Our expectation is that, 
$\delta N$
is identical to $\delta {\cal N}$ 
in the long wavelength limit in general. 
In fact, it was shown in Ref.~\cite{ST} that this is indeed
the case if $\eta_{\rm\, f}$ also lies in the scalar dominant phase. 
We plan to discuss this issue 
in the more general cases in our future publication~\cite{future}.
Here in this paper we simply assume $\delta {\cal N}\approx \delta N$ 
in the long wavelength limit. 
Notice that the meaning of this approximate equality is an 
identity relation which holds for any $\bbvp$ and $\bbvp'$. 
Hence, $W(\bbn, \bbvp)\approx W(\bbn_B,\bbvp)$ is also to 
be understood in the same way. 
Therefore we conclude $\bbn_B\approx\bbn$. 
In fact, it has been shown in our previous paper~\cite{ref4} 
that $\bbn_B$ defined above is a solution of Eq.~(\ref{flatEq}) 
in the long wavelength limit. 
Thus, in the case that the convergence of trajectories occurs 
after the scalar dominant phase ends, 
the boundary conditions for $\bbn$ are given by  
$\bbn=\bbn_B(\eta_{\rm\, i})$ and $ \bbn'=\bbn'_B(\eta_{\rm\, i})$ 
instead of Eq.~(\ref{basicbc}).

It would be worth mentioning that $\bbn$ used here 
is a full decaying mode solution obtained without assuming the 
long wavelength limit, 
while in Ref.~\cite{ref4} we used
a super-horizon decaying mode
solution, which is a solution for Eq.~(\ref{flatEq}) with $k^2 \to 0$. 
They are identical only in the long wavelength limit. 
Here we make full advantage of the constancy of the 
Wronskian $W(\bbn, \bbvp)$. 
In contrast, in Ref.~\cite{ref4} 
we discussed the evolution of the super-horizon Wronskian, 
which is defined by $W(\bbn, \bbvp)$ with $\bbn$ replaced by 
the super-horizon decaying mode solution. 

\section{Extended general slow-roll formula}
\label{extended}

Even if we use the Wronskian method, we still need to 
solve $\bbn$ backward for each $k$. However, we can avoid 
solving $\bbn$ for each value of $k$ in the long wavelength limit.  
In this limit, we can find the solution for $\bbn$ 
in the series expansion with respect to $k^2$. 
Of course,
this approximation does not hold when the wavelength becomes  
shorter than the horizon scale. For the evolution before the 
horizon crossing, one may use the slow-roll approximation or 
the more general formula called general slow-roll developed by 
Stewart\cite{ref3}. By using these approximations, 
we can evaluate 
the evolution of mode functions to a large extent analytically.
 
However, in the original slow-roll or general slow-roll formula~\cite{ref3,ref4},
one assumes no contribution on the super-horizon scales
\footnote{
This is roughly correct, but not exactly correct. More precisely, it assumes general slow-roll approximation is valid 
until superhorizon effects disappear.  
}
,
i.e., in these formulae, we cannot take into account the super-horizon effects
 pointed out in Ref.~\cite{ref2}.

Then, in order to take account of the super-horizon effects, 
one may naturally think of matching two approximation 
at an appropriate time $\eta=\eta_*$ 
where the mode is already well outside the 
horizon scale but the slow-roll or the general slow-roll 
conditions are still maintained. However, the matching does 
not seem to be so trivial since the solutions in the long 
wavelength approximation look quite different from those 
in the slow-roll or the general slow-roll approximation.

As an application of the Wronskian method, we consider this 
matching problem. The goal of this section is to obtain an 
improved general slow-roll formula. 
Since the Wronskian $W$ is constant in time, 
it is allowed to be evaluated at $\eta=\eta_*$. 
If we know both $\bbvp$ and $\bbn$ at $\eta=\eta_*$, 
what we have to do is simply to compute $W(\bbn,\bbvp)$ there. 
Then we can avoid the messy computation for 
explicitly matching solutions 
obtained in the two different approximation schemes term 
by term.  
To obtain $\bbvp$ and $\bbn$ at $\eta=\eta_*$, 
we evolve $\bbvp$ in the forward time direction 
by using the general slow-roll approximation, while 
$\bbn$ in the backward direction by using the long wavelength 
approximation.

\subsection{General slow-roll expansion}
\label{GSRE}
 
In the single field case,  under the general slow-roll condition,
slow-roll parameters, $\delta \equiv \ddot{\phi}/H\dot{\phi}$, 
and $\dot{\delta}/H$ can be the same order but both are assumed to be 
small. Here we used the dot as differentiation with respect to the cosmological time $t$ ($dt = ad\eta$)
and $H = \dot{a}/a$.
In this case $\bP$, which 
is $P^\alpha_{~\beta}(\eta)$ in matrix notation, can be shown to be small.
Here we mean that ${max}(P^{\alpha}_{~\beta})$ is small by
"the general slow-roll condition" as a simple 
generalization to the multi-component case.
In general, not all components of the multi-component scalar field 
are necessarily nearly massless.
 Hence, the assumption that ${max}(P^{\alpha}_{~\beta})$ is small is not 
satisfactory. We plan to propose an improved treatment in our 
future publication. 
We write
the solution solved in the forward direction in 
an expansion with respect to $\bP$ as 
\begin{equation} 
 \bbvp=\Delta\bbvp_{(0)}+\Delta\bbvp_{(1)}+\cdots,
 \label{bphi} 
\end{equation}
where $\Delta\bbvp_{(p)}$ is the correction of $O(\bP^{p})$. We also 
introduce variables
\begin{equation}
\bbvp_{(p)} \equiv\sum_{j=0}^p\Delta\bbvp_{(j)}.
\label{bphip}
\end{equation}
They satisfy the equation 
\begin{equation}
\left[\partial_\eta^2 + Q(\eta)+k^2-\bP\right]\bbvp_{(p)}
   = - \bP \Delta \bbvp_{(p)}.  
\end{equation}
Using the Green's function method, the first order correction with respect to $\bP$ is obtained as 
\begin{equation} 
 \Delta\bbvp_{(1)}= \Pi \left[
           u_0(\eta)\int_{-\infty}^{\eta} 
             {d\eta'} u_0^{*}(\eta')
                     \bP(\eta') \bbvp_{(0)}(\eta')
          -u_0^*(\eta)\int_{-\infty}^{\eta} 
             {d\eta'} u_0(\eta')
                     \bP(\eta') \bbvp_{(0)}(\eta')
               \right], 
\end{equation}
where $\Pi=W(u_0,u_0^*)^{-1}$, 
and $u_0(\eta)$ is a solution of 
\begin{equation}
(\partial_\eta^2 + Q(\eta)+k^2)u_0=0.
\label{u0}
\end{equation}

\subsection{$k^2$ expansion}
The backward solution obtained by setting the final 
condition is obtained as an expansion with respect to $k^2$
as
\begin{equation} 
 \bbn=\Delta \bbn^{(0)}+\Delta\bbn^{(1)}+\cdots,
\label{bn} 
\end{equation}
where $\Delta\bbn^{(q)}$ is $O(k^{2q})$. We also 
introduce 
\begin{equation}
\bbn^{(q)} \equiv \sum_{i=0}^q \Delta\bbn^{(i)}.
\label{bnq}
\end{equation}
They satisfy
\begin{equation}
(\partial_\eta^2 + Q(\eta)+k^2-\bP)\bbn^{(q)}
   = k^2 \Delta \bbn^{(q)}. 
\end{equation}

\subsection{Evaluation of Wronskian}
\label{evaluation}
To the first order in $\bP$ and to the $q$-th order in $k^2$, 
we have
\begin{eqnarray}
 {\cal R}_{(1)}^{(q)}&\equiv&  W(\bbn^{(q)}, \bbvp_{(1)})|_{\eta_*}
  = {\bbn^{(q)}}' \cdot \bbvp_{(1)}
       -\bbn^{(q)} \cdot {\bbvp_{(1)}}' \cr
 & = & W(\bbn^{(q)}, \bbvp_{(0)})|_{\eta_*}+
       \Pi\biggl[
         W(\bbn^{(q)}, u_0)|_{\eta_*}
       \int^{\eta_*}_{\etain} d\eta' 
       u^*_0(\eta') \bP(\eta') \bbvp_{(0)}(\eta')\cr
 &&\qquad\qquad\qquad\qquad\qquad
       -  W(\bbn^{(q)}, u^*_0)|_{\eta_*}
       \int^{\eta_*}_{\etain} d\eta' 
       u_0(\eta') \bP(\eta') \bbvp_{(0)}(\eta')
        \biggr]. 
\label{wronskian}
\end{eqnarray}

It is interesting to see how the above formula is systematically 
cast into the form 
appropriate for comparison with the standard slow-roll formula,~(\ref{standard}). 
The above formula still contains an ambiguous choice of the matching 
time $\eta_*$. Although the expression is 
independent of the choice of $\eta_*$, 
we cannot simply replace it with $\eta_k$, which is the time 
of the horizon crossing since the long wavelength approximation 
is not valid at $\eta=\eta_k$. 
The appropriate expression is 
\begin{eqnarray}
 {\cal R}_{(1)}^{(q)}&\equiv&  
    W(\bbn, \bbvp_{(0)})^{(q)}|_{\eta_k}\nonumber\\
    &&+\Biggl[\Pi\Biggl\{
         W(\bbn, u_0)
       \int^{\eta_*}_{\etain} d\eta' 
       \bP(\eta') \left[
             \tilde u^*_0(\eta') \tilde \bbvp_{(0)}(\eta')
            -\theta(\eta'-\eta_k) 
              u^*_0(\eta') \bbvp_{(0)}(\eta')\right]
            \cr
         &&\qquad\qquad
       -W(\bbn, u_0^*)
       \int^{\eta_*}_{\etain} d\eta' 
       \bP(\eta') \left[
            \tilde u_0(\eta') \tilde \bbvp_{(0)}(\eta')
            -\theta(\eta'-\eta_k) 
              u_0(\eta') \bbvp_{(0)}(\eta')\right]
            \Biggr\}
        \Biggr]^{(q)}_{\eta_k}. \nonumber\\
\label{multifirststep}
\end{eqnarray}
Here $f^{(q)}$ means the truncation of the function $f$ 
at $O(k^{2q})$, while the function with tilde $\tilde f$ means 
that we keep the original form of the function $f$ without 
expanding it in powers of $k^2$, 
namely, $\tilde f^{(q)}=f$.
This expression still contains $\eta_*$ but the 
integrand takes the form of 
$f-\theta (\eta-\eta_k)f^{(q)}$. 
Therefore all the terms up to $O(k^{2q})$ cancel, 
and hence the integrand 
shows improved fall off at $\eta\to 0$. 
As a result, even if fall off of $\bP$ in the limit $\eta\to 0$ 
is not very fast, 
we may be allowed to set $\eta_*$ to 0.
In this sense, the independence from the choice of $\eta_*$ 
is more manifest in Eq.~(\ref{multifirststep}). 

Now we will show that Eq.(\ref{multifirststep}) is equivalent to
Eq.(\ref{wronskian}). 
From Eqs.~(\ref{simplifiedequation}) and (\ref{u0}), we can easily get 
\begin{equation}
\left[\partial_\eta W(\bbn, u_0)\right]^{(q)}
  =\bP [\bbn u_0]^{(q)}. 
\label{eq11}
\end{equation}
Thus, we can show that 
\begin{eqnarray}
\partial_{\eta_k} {\cal R}_{(1)}^{(q)}
   &=&\left[\bbn\cdot \bP \bbvp_{(0)}\right]^{(q)}_{\eta_k} \cr\cr
   &&\quad + 
    \Biggl[ \Pi \Biggl\{ 
      \left[\bbn u_0 \right] \bP
       \int_{\etain}^{\eta_*} d\eta'
       \bP(\eta') \left[
            \tilde u^*_0(\eta') \tilde \bbvp_{(0)}(\eta')
            -\theta(\eta'-\eta_k) 
              u^*_0(\eta') \bbvp_{(0)}(\eta')\right]           
\cr\cr
   &&\qquad\qquad\qquad\qquad\qquad+
         \left[\bbn' u_0-\bbn  {u_0}'\right]
           \bP \left[u_0^* \bbvp_{(0)}\right]
\cr\cr        
    &&\qquad\qquad -\left[\bbn u^*_0\right] \bP 
       \int^{\eta_*}_{\etain} d\eta' 
       \bP(\eta') \left[
             \tilde u_0(\eta') \tilde \bbvp_{(0)}(\eta')
            -\theta(\eta'-\eta_k)u_0(\eta') \bbvp_{(0)}(\eta')\right]
            \cr\cr
   &&\qquad\qquad\qquad\qquad\qquad
      -\left[\bbn' u^*_0-\bbn {u^*_0}'\right]
           \bP \left[u_0 \bbvp_{(0)} \right]
       \Biggr\}\Biggr]^{(q)}_{\eta_k}~.
\label{eq12}
\end{eqnarray}
The first and the third lines in the curly brackets are 
second order in $\bP$. The second and the fourth terms inside the 
curly brackets are combined to give 
\[
\Biggl[\Pi W(u_0^*, u_0) \left[\bbn\cdot \bP \bbvp_{(0)}\right]\Biggr]^{(q)}_{\eta_k}
=-\left[\bbn\cdot \bP \bbvp_{(0)}\right]^{(q)}_{\eta_k}, 
\]
which cancels the first term on the right 
hand side, and hence $\partial_{\eta_k}{\cal R}_{(1)}^{(q)}=O(\bP^2)$.
 
Thus ${\cal R}_1^{(q)}$ is independent of the choice of $\eta_k$ to the first order in $\bP$. 
Setting $\eta_k\to \eta_*$ in ${\cal R}_1^{(q)}$, we find 
\begin{equation}
 W(\bbvp_{(1)},\bbn^{(q)})|_{\eta_*}={\cal R}_1^{(q)}+O(\bP^2,k^{2q+2}). 
\label{relation}
\end{equation}

\subsection{Extension to the higher order in general slow-roll expansion}

The extension to the higher order in $\bP$ is rather straight forward 
in our formulation. 
This clearly demonstrates the advantage 
of our systematic derivation. 
We introduce notations 
\begin{eqnarray}
\Delta{{\cal R}}_{(0)}^{(q)}  &\equiv&  \bigl[W(\bbn, \bbvp_{(0)})\bigr]^{(q)}_{\eta_k} ,\\
\hat{{\cal R}}[\bbvp] &\equiv&  \Pi \Biggl\{
         W(\bbn, u_0)
       \int^{\eta_*}_{\etain} d\eta' 
       \bP(\eta') \left[
             \tilde u^*_0(\eta') \tilde \bbvp(\eta')
            -\theta(\eta'-\eta_k) 
              u^*_0(\eta') \bbvp(\eta')\right]
             \cr
         &&\qquad
       -W(\bbn, u_0^*)
       \int^{\eta_*}_{\etain} d\eta' 
       \bP(\eta') \left[
            \tilde u_0(\eta') \tilde \bbvp(\eta')
            -\theta(\eta'-\eta_k)
              u_0(\eta') \bbvp(\eta')\right]
            \Biggr\}_{\eta_k}. \nonumber\\
\label{eq18}
\end{eqnarray}
In the similar way as we did in the above Eqs.~(\ref{eq11}) and
(\ref{eq12}), we obtain 
\begin{eqnarray}
\partial_{\eta_k} \Delta {\cal R}_{(0)}^{(q)} &=& [\bbn \cdot \bP
 \bbvp_{(0)}]^{(q)}_{\eta_{k}} , 
\cr
\partial_{\eta_k} \hat{\cal R}[\bbvp]^{(q)} &=& - [\bbn \cdot \bP \bbvp]^{(q)}_{\eta_{k}}
                           + \left[ \bbn \bP \Delta \hat \bbvp[\bbvp]\right]^{(q)}_{\eta_{k}}
                           +\left[\frac{\delta \hat{\cal R}[\bbvp]}{\delta
			   \bbvp}
			   \partial_{\eta_k}\bbvp\right]^{(q)}_{\eta_k}, 
\label{eq19}
\end{eqnarray}
where 
\begin{eqnarray}
\Delta \hat \bbvp[\bbvp] (\eta) &\equiv& \Pi\Biggl\{ u_0(\eta)
       \int^{\eta_*}_{\etain} d\eta'
       \bP(\eta') \left[
             \tilde u^*_0(\eta') \tilde \bbvp(\eta')
            -\theta(\eta'-\eta) 
              u^*_0(\eta') \bbvp(\eta')\right]
            \cr
         &&\qquad\quad
       -u_0^*(\eta)
       \int^{\eta_*}_{\etain} d\eta' 
       \bP(\eta') \left[
            \tilde u_0(\eta') \tilde \bbvp(\eta')
            -\theta(\eta'-\eta) 
                          u_0(\eta') \bbvp(\eta')\right]
            \Biggr\}.\nonumber\\
\end{eqnarray}
As a natural extension of Eq.~
(\ref{multifirststep}), we expect that 
${\cal R}$ including higher order corrections up to $O(\bP^p,k^{2q})$ 
will be given by 
\begin{eqnarray}
{{\cal R}}_{(p)}^{(q)} =&& \sum_{j=0}^p \Delta {{\cal R}}_{(j)}^{(q)}, 
\end{eqnarray}
with 
\begin{eqnarray}
&& \Delta  {{\cal R}}_{(1)}^{(q)} \equiv \hat{{\cal R}}[\bbvp_{(0)}]^{(q)}, 
\qquad
\Delta  {{\cal R}}_{(2)}^{(q)} \equiv \hat{{\cal R}}\left[ \Delta
				       \hat{\bbvp}[\bbvp_{(0)}]\right]^{(q)}, 
\cr\cr 
&&\Delta  {{\cal R}}_{(3)}^{(q)} \equiv \hat{{\cal R}}\left[
          \Delta\hat{\bbvp}[\Delta\hat{\bbvp}[\bbvp_{(0)}]]
                      \right]^{(q)}, 
\qquad \cdots. 
\end{eqnarray}
In fact, using Eqs.~(\ref{eq18}) and (\ref{eq19}), we can show that 
\begin{eqnarray}
\partial_{\eta_k} {\cal R}_{(p)}^{(q)} & = & \partial_{\eta_{k}} \biggl(\Delta {\cal R}_{(0)}^{(q)}
 +\hat{\cal R}[\bbvp_{(0)}]^{(q)} + \hat{\cal R}\left[ \Delta
 \hat{\bbvp}[\bbvp_{(0)}]\right]^{(q)} +\cdots \biggr) \nonumber\\
&=&  [\bbn \cdot \bP \bbvp_{(0)}]^{(q)}_{\eta_{k}} - [\bbn \cdot \bP
 \bbvp_{(0)}]^{(q)}_{\eta_{k}}
       + \left[\bbn \bP \Delta \hat \bbvp[\bbvp_{(0)}]\right]^{(q)}_{\eta_{k}}
\cr\cr
     &&- [\bbn \cdot \bP  \Delta \hat{\bbvp}[\bbvp_{(0)}]]^{(q)}_{\eta_{k}} 
    + \left[ \bbn \bP \Delta \hat \bbvp\left[ \Delta
   \hat{\bbvp}[\bbvp_{(0)}]\right]\right]^{(q)}_{\eta_{k}}
+\cdots \cr\cr
&=&
O(\bP^{p+1}), 
\label{eq21}
\end{eqnarray}
where we used the fact that the arguments of $\hat {\cal R}$, such as 
$\bbvp_{(0)}$ and $ \Delta \hat{\bbvp}[\bbvp_{(0)}]$, are
independent of $\eta_k$. 
Eq.~(\ref{eq21}) shows that ${\cal R}_{(p)}^{(q)}$ is independent of the choice 
of $\eta_k$ to the $p$-th order in $\bP$. 
Setting $\eta_k\to \eta_*$ in ${\cal R}_{(p)}^{(q)}$, we find 
\begin{equation}
 W(\bbvp_{(p)},\bbn^{(q)})|_{\eta_*}={\cal R}_{(p)}^{(q)}+O(\bP^{p+1},k^{2q+2}). 
\end{equation}\\

\subsection{Power spectrum}
The perturbation $\varphi_{(0)}^a$ is to be treated as a 
quantum operator, and it will be expanded as 
\begin{equation}
\varphi_{(0)}^\alpha=u_0 a^\alpha +u^*_0 a^{\alpha\dag}. 
\end{equation}
Creation and annihilation operators have expectation value
\begin{equation}
\langle a^\alpha a^{\beta\dag}\rangle =\delta^{\alpha\beta}.
\end{equation}
The expectation values of the 
other combinations of two creation and annihilation operators 
are zero. 

Then, to the first order in $\bP$ and to the $q$-th order in $k^2$, the
expectation value of $|{\cal R}|^2$ 
will be evaluated as
\begin{eqnarray}
{\cal P}_{{\cal R}} \sim \langle|{{\cal R}}|^2\rangle_{(1)}^{(q)}
=&\Biggl[& |\bPhi|^2 \cr\cr
 && +4i \Pi \int_{\etain}^{\eta_*}d\eta' 
\Biggl\{{\rm Im}\left[\tilde u_0^\ast(\eta')\bPhi \right] \cdot \bP(\eta')
        {\rm Re}\left[\tilde u_0^\ast(\eta')\bPhi \right]\cr\cr
                &&\qquad\qquad-\theta(\eta'-\eta_k) {\rm Im} \left[ u_0^\ast(\eta')\bPhi \right] \cdot \bP(\eta')
        {\rm Re}\left[u_0^\ast(\eta')\bPhi \right] \Biggr\} \Biggr]^{(q)}
       +{\cal O}(\bP^2),\nonumber\\
\label{PS}
\end{eqnarray}
with
\begin{equation}
 \bPhi\equiv [W(\bbn,u_0)]_{\eta_k}.
\end{equation}
Here we should remind that 
$\Pi$ is a pure imaginary number.

In Sec.~\ref{GSRE}, in order to obtain the solution solved in the forward direction using general slow-roll 
approximation,
we have assumed that all components of the multi-component scalar field are nearly massless.
However, in Eq.~(\ref{PS}) the first order term in $\bP$ takes the form of ${\rm Im}\left[\tilde u_0^\ast(\eta)\bPhi \right] \cdot \bP(\eta){\rm Re}\left[\tilde u_0^\ast(\eta)\bPhi \right]$.
Therefore, in this formula there is no need to assume all components are nearly massless.
We only need to assume that the component of $\bP$ connected to $\bPhi$ is small.
This seems a natural result, because
basically the advantage of Wronskian formulation is that we only need to solve a single mode, as we mentioned before.
We will prove the validity of this argument rigorously in our future paper~\cite{future}.
       
\section{Single Field}
\label{secsingle}
Here, we consider the single scalar field case. 
In this case, we can easily obtain more explicit solutions 
described in an expansion with respect to $k^2$ and $g$, which is correspond to $\bP$ in Sec.~\ref{extended}, respectively.
Furthermore, it is known that the usual approach that uses ${\cal R}_c$ 
as a perturbation variable apparently breaks down when $\phi'=0$. 
At that point special treatment is necessary to justify 
the evaluation of perturbation by means of ${\cal R}_c$~\cite{Seto}.  
What we evaluate here is a similar quantity 
${\cal R}_c(\eta_{f})\equiv W$ but its argument 
is slightly different from 
${\cal R}_c(\eta)$. Unless we choose $\eta_{\rm\, f}$ so that
$\phi'(\eta_{\rm\, f})=0$, 
${\cal R}_c(\eta_{\rm\, f})$ must be free from any singular behavior. 
This fact will become manifest below. 

\subsection{Extended general slow-roll formula in the single field case}
In the case of single field, the equation 
for $\delta\phi_F$ (\ref{flatEq})
reduces to 
\begin{eqnarray}
\left(\partial_\eta^2+Q(\eta)+k^2\right)\varphi(\eta) 
= \frac{g(\eta)}{\eta^2} \varphi(\eta), 
\end{eqnarray}
with
\begin{equation}
Q(\eta) = -{2\over \eta^2},\qquad 
g(\eta) = \eta^2\frac{z''}{z} - 2, 
\label{perturbation}
\end{equation}
where $z = a\phi'/{\cal H}$. 
 
The solution in the long wavelength approximation $n^{(i)}$ satisfies
\begin{eqnarray}
\left(\partial_\eta^2 - \frac{z''}{z}\right) n^{(0)} = 0~,~~ 
\left(\partial_\eta^2 -\frac{z''}{z}\right) n^{(q)}
   = -k^2 n^{(q-1)}~. 
\end{eqnarray}
The boundary conditions (\ref{basicbc}) at $\eta=\eta_{\rm end}$ reduce to 
\begin{eqnarray} 
n \equiv 0~,~~ 
n' \equiv -1/z~, 
\label{final}
\end{eqnarray}
where $\eta_{\rm end}$ is taken as a time when scalar dominant phase ends.

To the first order in $k^2$, we have
\begin{eqnarray}
n^{(1)} &=& n^{(0)}
         +k^2\Delta n^{(0)}~ \nonumber\\
 &=& -z(\eta)\int_{\eta_{\rm end}}^{\eta}\frac{d\eta'}{z^2(\eta')}+k^2
z(\eta)\int_{\eta_{\rm end}}^{\eta}d\eta'\frac{I(\eta')}{z^2(\eta')}~,\nonumber\\
I(\eta) &\equiv& \int_{\eta _{\rm end}}^{\eta}d\eta' z^2(\eta')\int_{\eta_{\rm end}}^{\eta'}\frac{d\eta''}{z^2(\eta'')}~.
\label{n}
\end{eqnarray}

As for the $g$-expansion, now 
the lowest order mode function $u_0$, which is a solution 
of 
\begin{eqnarray}
\left(\partial^2_\eta+k^2-\frac{2}{\eta^2}\right)u_0 = 0, 
\end{eqnarray}
is explicitly given by 
\begin{eqnarray}
u_0 = -\frac{i}{\sqrt{2k^3}}
  \left(ik+\frac{1}{\eta}\right)e^{-ik\eta}~.
\label{varphi}
\end{eqnarray}
Here we have chosen $u_0$ such that it corresponds to 
the positive frequency function of the Bunch-Davies vacuum, \mbox{i.e.}
\begin{equation}
u_0 \to \frac{1}{\sqrt{2k}}e^{-ik\eta}~~~{\rm for}~~~ \eta \to - \infty.
\end{equation}

In the present single field case, 
using Eqs.~(\ref{n}) and (\ref{varphi}), 
the power spectrum of ${\cal R}_c$,
which is given in Eq.~(\ref{PS}), to 
 the first order in $g$ and $k^2$ 
is more explicitly written as 
\begin{eqnarray}
{\cal P}_{{\cal R}_c} &\simeq& 
\left(\alpha^2 + 2k^2\alpha\beta\right)_{\eta_k}
\left[1-{2 \over 3}
\int^{\eta_*}_{\etain} 
d\eta'\frac{g(\eta')}{\eta'}
{\cal F}(\eta')\right]\nonumber\\
&&\qquad\qquad\qquad
+2\left[\alpha\gamma+k^2(\alpha\sigma+\beta\gamma)\right]_{\eta_k}
\int^{\eta_*}_{\etain} 
d\eta'\frac{g(\eta')}{{\eta'}^4}
{\cal G}(\eta')~,
\nonumber\\
\label{power}
\end{eqnarray}
where
\begin{eqnarray}
\alpha(\eta) &\equiv& -\frac{1}{\eta^2}
\left(\eta{n^{(0)}}\right)' 
~,\qquad\qquad
\beta(\eta) \equiv -\frac{\eta^2}{2}\left({n^{(0)} \over \eta}\right)'
-\frac{1}{\eta^2}\left(\eta \Delta n^{(1)}\right)'
~,
\nonumber\\
\gamma(\eta) &\equiv& -\frac{\eta^4}{3}\left({n^{(0)} \over \eta^2}\right)'~,\qquad\qquad
\sigma(\eta) \equiv \frac{\eta^8}{30}\left({n^{(0)} \over \eta^4}\right)'
-\frac{\eta^4}{3}\left({\Delta n^{(1)}\over \eta^2}\right)'~, \nonumber\\
{\cal F}(\eta) &\equiv& {\rm Re}[F(\eta)]-\theta(\eta-\eta_k){{\rm Re}[F(\eta)]}^{(1)}~,
\qquad{\cal G}(\eta) \equiv {\rm Re}[G(\eta)]-\theta(\eta-\eta_k){{\rm Re}[G(\eta)]}^{(1)},\nonumber\\
\end{eqnarray}
and where
\begin{eqnarray}
F(\eta) &\equiv& -\frac{3i}{\eta}\left(u_{0}u_{0}^{\ast}+u_{0}u_{0}\right)
=1+\frac{2}{5}k^{2}\eta ^{2}+\cdots~,\nonumber\\
G(\eta) &\equiv& k^3 \eta^2
 \left(u_{0}u_{0}^{\ast}-u_{0}u_{0}\right)=1+k^{2}\eta^{2}+\cdots~. 
\end{eqnarray}
Then, the asymptotic behavior of ${\cal F}(\eta)$ and ${\cal G}(\eta)$ is 
\begin{eqnarray}
\lim_{k\eta \to -\infty}{\cal F}(\eta)&=&{3\over k\eta}\sin(-k\eta)\cos(-k\eta),\\
\lim_{k\eta \to -\infty}{\cal G}(\eta)&=&{(-k\eta)^2\sin^2(-k\eta)},\\
\lim_{k\eta \to 0}{\cal F}(\eta)&=& -{6 \over 35}(-k\eta)^4+O\left((-k\eta)^6\right),\\
\lim_{k\eta \to 0}{\cal G}(\eta)&=& -{1 \over 9}(-k\eta)^6+O\left((-k\eta)^8\right).
\end{eqnarray}

\begin{figure}[htbp]
 \begin{center}
  \includegraphics[keepaspectratio=true,height=50mm]{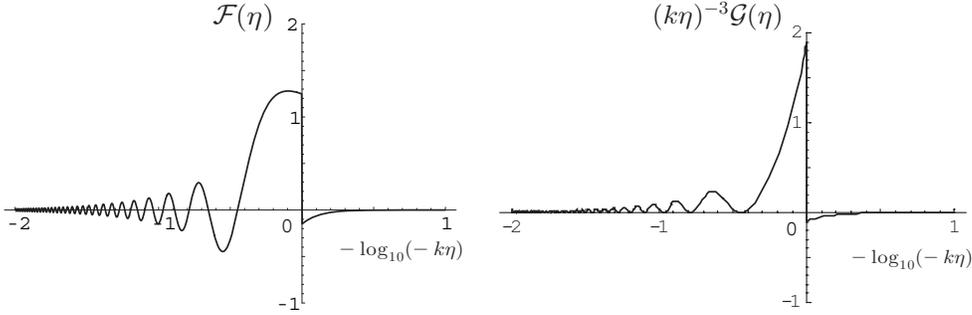}
  \end{center}
  \caption{The window functions ${\cal F}(\eta)$ and $(k\eta)^{-3}{\cal G}(\eta)$ as functions of $-\log (-k\eta)$.}
  \label{fig1}
\end{figure}

Here we show the plots of 
the window functions ${\cal F}(\eta)$ and 
$ {\cal G}(\eta)/{\eta}^3$ in FIG.~\ref{fig1}.
FIG.~\ref{fig1} tells that 
we have only to consider the contribution from the neighborhood
of the horizon crossing time. 
Moreover, from these window functions' behavior, we can easily find that
we may be allowed to set $\eta_*$ to 0 even if fall-off of $g(\eta)$ in the limit $\eta\to 0$ 
is not very fast, as we mentioned in Sec. \ref{evaluation}.

\subsection{$\phi'=0$}
When $\phi' = 0$, 
it seems that the solution $n^{(1)}$ diverges because $z$ vanishes, 
but we will see below that in fact it does not.  

Using integration by parts, we can rewrite the integral $\int
d\eta'/z^2(\eta')$ which appears in the Eq.~(\ref{n}) as
\begin{eqnarray}
n^{(0)}(\eta) = -z(\eta)\int_{\eta _{\rm end}}^{\eta} \frac{d \eta'}{z^{2}(\eta')}
&=&z(\eta)\int_{\eta _{\rm end}}^{\eta}d \eta'
\left\{\frac{1}{z'(\eta')}\frac{d}{d \eta'}\left(\frac{1}{z(\eta')}\right)\right\} \nonumber\\
&=&-z(\eta) \left[ -\frac{1}{z'(\eta')z(\eta')}\right]^{\eta}_{\eta _{\rm end}}
-z(\eta)\int_{\eta _{\rm end}}^{\eta}d \eta'
\left\{\frac{1}{z(\eta')}\frac{d}{d \eta'}\left(\frac{1}{z'(\eta')}\right)\right\} \nonumber\\
&=&\frac{1}{z'(\eta)}
-\frac{z(\eta)}{z(\eta_{\rm end})z'(\eta_{\rm end})}
+z(\eta)\int_{\eta _{\rm end}}^{\eta}d \eta' \left(\frac{z''(\eta')}{z(\eta')z'^{2}(\eta')}\right). \label{inflatonstop1}
\end {eqnarray}
In a similar way, we also have
\begin{eqnarray}
\Delta n^{(0)}&=& z(\eta)\int_{\eta_{\rm end}}^{\eta}d\eta'\frac{I(\eta')}{z^2(\eta')}\nonumber\\
&=& -\frac{1}{z'(\eta)}+\frac{z(\eta)}{z(\eta_{\rm end})z'(\eta_{\rm end})} \nonumber\\
&&-z(\eta)\int_{\eta _{\rm end}}^{\eta}d \eta'\frac{z''(\eta')}{z(\eta')z'^{2}(\eta')}
\int_{\eta _{\rm end}}^{\eta'}d \eta'' z(\eta'')n^{(0)}(\eta'')
+z(\eta)\int_{\eta _{\rm end}}^{\eta}d \eta'\frac{n^{(0)}(\eta')}{z'(\eta')}. \nonumber\\
\end{eqnarray}
Here, $z'$ and $z''/z$ can be rewritten explicitly as 
\begin{eqnarray}
z' &=& a^2\left(\frac{{\phi'}^3}{2{\cal H}^2}- 2\phi'- \frac{a^2V_{,\phi}}{{\cal H}}\right)~, \cr\cr
z''/z &=& 2{\cal H}^2\left(1- \frac{a^2V_{,\phi\phi}}{2{\cal H}^2}-\frac{a^2\phi'V_{,\phi}}{{\cal H}^3}
-\frac{11}{4}\frac{{\phi'}^2}{{\cal H}^2} + \frac{{\phi'}^4}{4{\cal H}^4}\right)~. 
\label{zdoubleprime}
\end{eqnarray}
Hence, we have $z' \neq 0$ and $z''/z \neq \infty$ when $\phi' = 0$,
as long as $V_{,\phi}\neq 0$.
\footnote{
We can easily consider such cases, for example, the oscillating
inflation proposed by Damour and Mukhanov~\cite{DM}.
} 
Therefore we find that $n^{(1)}$ doesn't diverge even if 
$\phi'$ vanishes.

Finally, it will be worth pointing out 
that the general slow-roll condition is even weaker than those shown in Ref.~\cite{ref3}.  
As noted earlier, under the general slow-roll condition,
the slow-roll parameters, $\delta \equiv \ddot{\phi}/H\dot{\phi}$, 
and $\dot{\delta}/H$ can be the same order but both are assumed to be 
small in Ref.~\cite{ref3}.
However, we can rewrite $g=\eta^2 z''/z -2$ by using 
the expression for $z''/z$ given in Eq.~(\ref{zdoubleprime}) and 
\begin{eqnarray}
\eta^2 &=& \frac{1}{{\cal H}^2}\left(1 + 2\epsilon +
			     O(\epsilon^2)\right),\nonumber\\
\label{phi}
\end{eqnarray}
where $\epsilon \equiv -\dot{H}/H^2$.
General slow roll assumes $|g| \ll 1$. 
When the background scalar field motion stops
($\dot{\phi} \simeq 0$), the slow-roll parameter $\delta$ diverges. 
Therefore, slow-roll conditions are violated at that point. 
However, from Eq.~(\ref{zdoubleprime}), we can easily find that 
the condition $|g|\ll 1$ is maintained as far as 
$V_{,\phi \phi}/{H^2} \ll 1$ even when $\dot{\phi} \simeq 0$, 
namely, $\dot\phi\simeq 0$ does not imply the breakdown of 
the general slow-roll condition.

\section{Summary}
\label{summary}

We have proposed a new method for a systematic derivation of formulas for 
the spectrum of the curvature perturbations from multi-component inflation. 
First we have noted that the Wronskian $W(\bbn,\bbvp)$ between
two solutions $\bbn$ and 
$\bbvp$ of the scalar field perturbation equation in flat slicing 
stays constant in time.
Using this fact, we have shown that the evaluation
of the curvature perturbations at the end of inflation reduces to
an easier problem of solving a single 
``decaying'' mode $\bbn$ backward in time from the end of inflation. 

We have also shown that, for any given perturbation $\bbvp$  
the Wronskian $W(\bbn, \bbvp)$ becomes identical to 
$\delta N$ in the long wavelength limit 
by choosing an appropriate boundary condition for $(\bbn, \bbn')$,
where $\delta N$ is the perturbation in the $e$-folding number 
along the phase space trajectory of the homogeneous
universe. More precisely, $\delta N$ is a perturbation in the
$e$-folding number from a given point in the phase space for
the homogenous universe to a point when the
trajectories converge to a unique one, with $(\bbvp,\bbvp')$ 
identified as the perturbation in the background phase space
of spatially homogeneous fields.

Hence, with the aid of the $\delta N$ formalism~\cite{SS,ST},
even if the phase space trajectories converge only 
after the scalar dominant phase ends,
we can still evaluate the final value of the curvature perturbation
by evaluating the Wronskian.
Yet an open question is whether the perturbation in the true 
$e$-folding number, $\delta {\cal N}$, 
is well approximated by $\delta N$, which is determined by 
the dynamics of the homogeneous 
universe for the case with general matter fields.
We plan to discuss this issue in our future publication~\cite{future}.

As a good example to show the efficiency of the new formulation 
using the Wronskian, 
we have presented a quick derivation of an improved general 
slow-roll formula, 
whose original form was obtained in Ref.~\cite{ref3}. 
In Ref.~\cite{ref3} or in the general slow-roll formulae given in 
Ref.\cite{ref4}, it was assumed 
that super-horizon effects are negligible. However, it was
pointed out in Ref.~\cite{ref2}
that there are cases in which they may not be negligible at all.

The improvement that has been incoorporated for the first time 
in this paper is to use a systematic long-wavelength 
expansion for the late time evolution after horizon crossing.
The matching of the two approximations, the general slow-roll
approximation and the long wavelength expansion, 
has been easily performed by using the constancy of the Wronskian. 
We have shown that the formula which contains 
higher order corrections to the general slow-roll approximation 
can be obtained systematically. 
We have also presented an explicit power spectrum formula for
the single field case. 
Furthermore, we have shown that our formula is 
valid even if the scalar field passes the point $\dot{\phi}=0$
where one of the slow-roll parameters, $\delta$, diverges 
($\delta\equiv\ddot{\phi}/H\dot{\phi}\to \infty$).
This fact that any special treatment is not required even if
 $\dot{\phi}=0$ occurs is a small but important advantage of
our formulation in comparison over previous treatments~\cite{Seto}.

\subsection*{Acknowledgements}
We would like to thank Takashi Nakamura for useful comments.
This work is supported in part by
ARCSEC funded by the Korea Science and Engineering Foundation and the
Korean Ministry of Science,
the KOSEF Grant (KOSEF R01-2005-000-10404-0),
an Erskine Fellowship of the University of Canterbury,
and JSPS Grant-in-Aid for Scientific Research, Nos.~14102004,
16740165 and 17340075.

\end{document}